\RequirePackage{fix-cm}
\documentclass[twocolumn,epjc3]{svjour3}

\smartqed  
\RequirePackage{graphicx}
\RequirePackage{mathptmx}      
\RequirePackage[numbers,sort&compress]{natbib}
\RequirePackage[colorlinks,citecolor=blue,urlcolor=blue,linkcolor=blue]{hyperref}
\usepackage{doi}
\usepackage{textcomp}
\usepackage{color}
\usepackage{amsmath}
\usepackage[utf8]{inputenc} 
\usepackage{comment}

\newcommand{\up}[1]{\mathrm{#1}}

\newcommand{\milli}{m}
\newcommand{\micro}{\mbox{$\mu$}}
\newcommand{\nano}{n}
\newcommand{\degree}{\mbox{$^\circ$}}
\newcommand{\squared}{\mbox{$^2$}}

\newcommand{\electronvolt}{eV}
\newcommand{\meter}{m}
\newcommand{\second}{s}
\newcommand{\tesla}{T}
\newcommand{\hertz}{Hz}
\newcommand{\volt}{V}
\newcommand{\mcps}{mcps}
\newcommand{\SI}[2]{\mbox{#1~#2}}

\journalname{Eur. Phys. J. C}

\begin{document}

\title{Time-Focusing Time-of-Flight, a new method to turn a MAC-E-filter into a quasi-differential spectrometer}


\author{A. Fulst\thanksref{e1,addr1}
        \and
        A. Lokhov\thanksref{addr1,addr2} 
        \and
        M. Fedkevych\thanksref{addr1,addr3,addr4} 
        \and
        N. Steinbrink\thanksref{addr1}
        \and
        C. Weinheimer\thanksref{addr1}
}

\thankstext{e1}{e-mail: alexanderfulst@uni-muenster.de}


\institute{Institut f\"ur Kernphysik, Westf\"alische Wilhelms-Universit\"at M\"unster, Wilhelm Klemm-Strasse 9, D-48149 M\"unster, Germany \label{addr1}
           \and
           Institute for Nuclear Research of Russian Academy of Sciences, 60th October Anniversary Prospect 7a, 117312 Moscow, Russia \label{addr2}
           \and
           \emph{Present Address:} Dipartimento di Fisica, Università di Genova, Genova, Italy\label{addr3}
           \and
           \emph{Present Address:} Istituto Nazionale di Fisica Nucleare (INFN), Sezione di Genova, Genova, Italy\label{addr4}
}

\date{Received: date / Accepted: date}

\maketitle

\begin{abstract}
Spectrometers based on the magnetic adiabatic collimation followed by an electrostatic filter (MAC-E-filter) principle combine high angular acceptance with an excellent energy resolution.
These features make MAC-E-filters very valuable for experiments where the kinetic energy of ions or electrons from rare processes has to be measured with utmost sensitivity and precision.
Examples are direct neutrino mass experiments like KATRIN which investigate the energy of electrons in the endpoint region of the tritium $\beta$-spectrum.
However, the MAC-E-filter is a very sharp energy high-pass filter but not a differential spectrometer.
To determine a spectral shape of a charged particle source, different electric retarding potentials have to be used sequentially, reducing the statistics.

In a previous work we have shown that the advantages of the standard MAC-E-filter can be combined with a measurement of the time-of-flight (TOF), allowing to determine spectral information over a certain energy range with one retarding potential only, with the corresponding gain in statistics.
This TOF method requires one to know the start time of the charged particles, which is not always possible.
Therefore, we propose a new method which does not require the determination of the start time and which we call "time-focusing Time-of-Flight" (tfTOF):
By applying a time dependent acceleration and deceleration potential at a subsequent MAC-E-filter, an energy dependent post-bunching of the charged particles is achieved.
\end{abstract}

\section{Introduction}
\label{sec:intro}
In electron or ion spectroscopy at low count rates there are often strong requests for excellent energy resolution in combination with a large solid angle acceptance.
An electrostatic retarding filter following adiabatic collimation (``MAC-E-filter'') nicely fulfills both requirements.
Originally invented for photoelectron spectroscopy \cite{Beamson1980}, it was reinvented for $\beta$-electron spectroscopy for experiments in search of the neutrino mass \cite{Lobashev1985,Picard1992} and applications to ion spectroscopy comprise precision recoil spectroscopy after $\beta$-decay \cite{Finlay2016,Baessler2008}.
The Karlsruhe Tritium Neutrino experiment KATRIN \cite{KATRIN2005,Aker2019} is even using a tandem of two MAC-E-filters in series to perform a neutrino mass search with \SI{0.2}{\electronvolt} sensitivity.
The larger of the two, the KATRIN main spectrometer with its length of \SI{24}{\meter} and its diameter of \SI{10}{\meter}, is providing an energy resolution of down to\SI{1}{\electronvolt} at \SI{18600}{\electronvolt} and an accepted solid angle of $0.7\pi$.

However, the advantages of high acceptance and excellent energy resolution of a MAC-E-filter come with the fact that it is a very sharp energy high-pass filter but not a differential spectrometer.
All charged particles above a certain energy threshold, defined by the retarding potential, are transmitted.
To determine their spectral shape, the count rates at different electric retarding potentials have to be measured sequentially, which reduces the statistics by the number of retarding potential set points.
Therefore, a huge gain in statistics or a reduction in measurement time could be achieved by combining the integral MAC-E-filter technology with a differential method.
One possibility would be to use only the lowest filter potential and determine the energy of the transmitted electrons (ions) with an ultra-high resolution detector.
Unfortunately, an array of cryo-bolometers as detector is not an obvious choice, since the MAC-E-filter requires a very strong magnetic field to guide the electrons (ions) to the detector and most high resolution read-outs of these detectors rely on superconducting sensor techniques that are not suitable for use in magnetic fields of Tesla strength.

Therefore, a new differential method is desired for the advancements of MAC-E-filters.
In a previous publication the possibility to measure the time-of-flight (TOF) of electrons (ions) through a MAC-E-filter to obtain an additional differential method has been investigated \cite{Steinbrink2013}.
It relies on the fact that the retarding potential of the MAC-E-filter increases the TOF of the electron (ion) dramatically and small differences in energy above threshold manifest themselves in significantly different TOF values.
It has been shown that for the neutrino mass measurement with a KATRIN-like experiment an improvement of about a factor 5 in the sensitivity to $m^2(\nu_e)$ is possible, if the TOF can be determined with a precision of \SI{50}{\nano\second}.
The TOF method at a MAC-E-filter 
has been successfully applied for the determination of the response function of the KATRIN experiment with electrons from a pulsed photoelectron source \cite{Aker2019}.
Here, the TOF is determined by the difference of the arrival time at the detector and the laser pulse of the photoelectron source.
Unfortunately, in many other cases, e.g.\ KATRIN's main purpose, the tritium $\beta$-electron spectroscopy, it is very difficult or even impossible to assess the start time of the electron (ion).

In this paper we present a modification of the TOF method which does not require the determination of the start time of the charged particle under investigation.
Instead, the electron (ion) beam is exposed to a periodically varying electrostatic potential after the energy filtering in the analyzing plane of the MAC-E-filter.
This can be performed in a dedicated hardware section equivalent to a spectrometer of the MAC-E-filter type functioning as delay line.
The aim is to apply the electric potential of the delay line with an appropriate waveform, frequency and amplitude in order to accelerate electrons (ions) which arrive later w.r.t.\ the pulsing phase in such a way that all electrons (ions) with a given energy of interest arrive at the same time at the detector.
This is somewhat similar to the known velocity focusing of mass spectrometers, where ions of the same mass are brought to a focus, regardless of their initial velocities.
Therefore, we call this new technique ``time-focusing Time-of-Flight'' (tfTOF).

We will present the idea in detail in the next sections.
Thereafter, we will discuss the prospects of applying the technique to the KATRIN experiment and present simulations to assess its potential sensitivity to the neutrino mass search.

\section{Magnetic Adiabatic Collimation with Electrostatic Filtering}
\label{sec:mac-e}
The MAC-E-filter technique is based on a simple way to analyze the kinetic energy of charged particles, namely letting them pass or reflecting them with an electrostatic potential.
Because a retarding potential $U$ acts as a high-pass filter on the longitudinal kinetic energy $E_\parallel$ only and not on the transversal component $E_\perp$, an energy conserving collimation is necessary to be able to analyze the total energy $E_\up{kin}=E_\perp+E_\parallel$ of a particle.
The MAC-E-filter therefore combines a high magnetic field on the order of several Tesla in the source and at the detector with a low magnetic field of \SI{\milli\tesla} strength or below in the analysis plane of the spectrometer where the retarding potential is maximal.
If the change of magnetic field during a cyclotron turn is small, i.e. $\left| \frac{1}{B}\frac{\up{d}B}{\up{d}t} \right| \ll\frac{\omega_c}{2\pi}$ with $\omega_c$ being the cyclotron frequency, the magnetic moment $\mu$ of the electron is a conserved quantity:
\begin{equation}
\label{eq:magnetic_moment}
\mu = \frac{E_\perp}{B} = const.
\end{equation}
From eq.~\eqref{eq:magnetic_moment} it follows that the transverse energy of an electron starting in a high magnetic field is converted into longitudinal energy in a region of low magnetic field, which can be analyzed by the electrostatic potential.
The energy resolution $\Delta E$ of such a filter is defined by the maximal transversal energy an electron can have in the analysis plane
\begin{equation}
\frac{\Delta E}{E}=\frac{B_\up{min}}{B_\up{max}}\ ,
\end{equation}
which is only depending on the ratio of the minimal and maximal magnetic field strengths.

In the case of KATRIN the design value for the maximal magnetic field is $B_\up{max}=\SI{6}{\tesla}$, while the minimal magnetic field is designed to be $B_\up{min}=\SI{0.3}{\milli\tesla}$, resulting in a sub-eV energy resolution of \SI{0.93}{\electronvolt} at the tritium endpoint $E_0 \approx \SI{18600}{\electronvolt}$.
In addition, the start magnetic field $B_0=\SI{3.6}{\tesla}$ is chosen to be lower than the maximal field, so that only electrons with an initial pitch angle $\theta_0$ towards the magnetic field of less than
\begin{equation}
\label{eq:max_start_angle}
\theta_\up{max}= \arcsin{\sqrt{\frac{B_0}{B_\up{max}}}}
\end{equation}
are able to pass the pinch magnetic field $B_\up{max}$.
Electrons with high initial angles are magnetically reflected in order to reject long flight paths inside the source, avoiding higher scattering probabilities in the tritium source and thus larger energy loss systematics.

\begin{figure}[hbt]
	\centering
	\includegraphics[width=\linewidth]{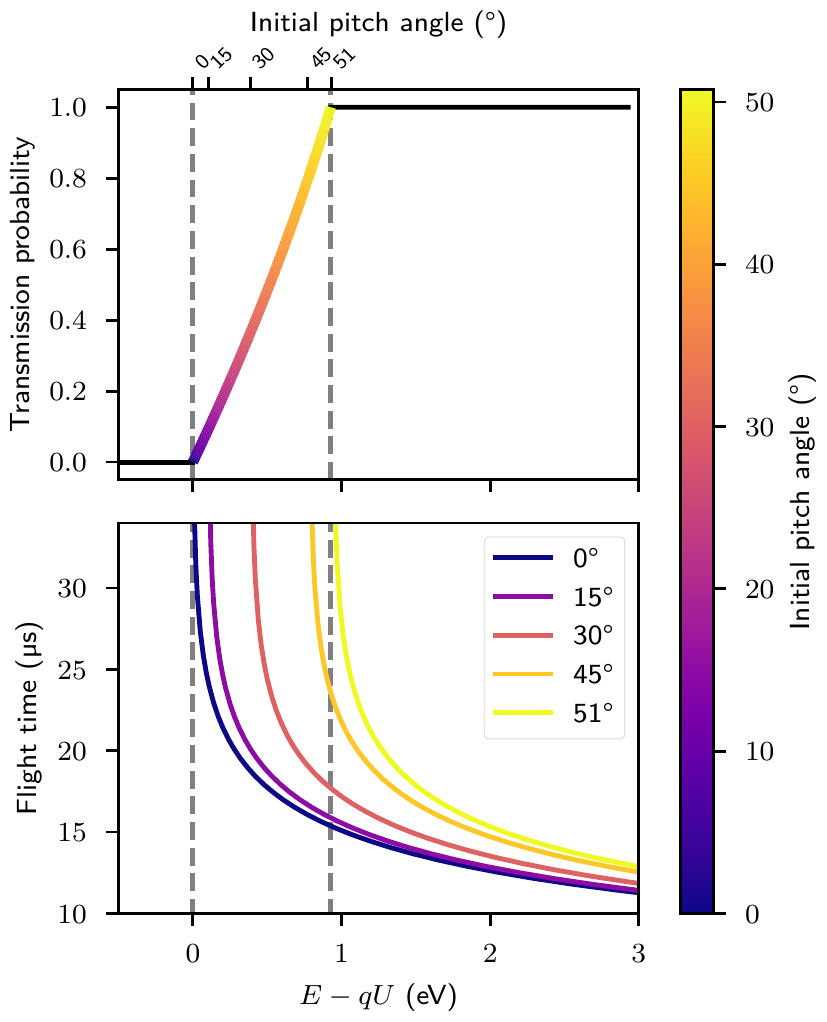}
	\caption{Transmission probability (top) and flight time (bottom) in dependence of the surplus energy and initial pitch angle (color code) in the case of KATRIN.
	The maximal transmitted angle grows from $\theta=\SI{0}{\degree}$ to $\theta=\SI{51}{\degree}$ as the energy of the electrons increases.
	The gray dashed lines indicate the energy resolution of the MAC-E-filter.
	While the flight times for electrons with the same energy but different angles converge for energies high above the retarding potential, they are distinct for lower values.}
	\label{fig:transmission}
\end{figure}

For an isotropic source of particles with kinetic energy $E$, the normalized transmission function $T(E,U)$ of the MAC-E-filter with a given retarding energy $qU$, with $q$ being the charge of the electron or ion, is then piece-wise defined to be:
\begin{equation}
\label{eq:transmission}
T(E,U)=
	\begin{cases}
		0 & \text{, }E\leq qU\\
		\frac{1-\sqrt{1-\frac{E-qU}{E}\frac{B_0}{B_\up{min}}}}{1-\sqrt{1-\frac{B_0}{B_\up{max}}}} & \text{, }qU < E < qU+\Delta E \\
		1 & \text{, }E\geq qU+\Delta E
	\end{cases}\,.
\end{equation}
\noindent The transmission function according to eq.~\eqref{eq:transmission} is shown in the upper part of figure~\ref{fig:transmission} for the KATRIN main spectrometer values $B_\up{min}=\SI{0.3}{\milli\tesla}$, $B_\up{max}=\SI{6}{\tesla}$, $B_0=\SI{3.6}{\tesla}$ and $qU=\SI{18560}{\electronvolt}$.

From the lower plot in figure~\ref{fig:transmission}, it can be seen as well that the time-of-flight highly depends on the surplus energy of the electron, because the initially fast electrons are slowed down significantly in the MAC-E-filter due to the retarding potential.
A TOF measurement thus can in principle be used to obtain the differential energy information from the electron \cite{Steinbrink2013}.
The challenging part of this method is however the determination of the start time of the electron.
In a situation with a continuously emitting source, the electron would have to be detected in front of the MAC-E-filter without disturbing its energy and angle excessively.

Because this problem remains unsolved as of yet \cite{Steinbrink2013,Steinbrink2018}, we decide for a different approach of obtaining energy information from the flight time.

\section{Principles of Time-Focusing Time-of-Flight}
\label{sec:tftof}
Before we discuss the time-focusing Time-of-Flight (tfTOF) method for a MAC-E-filter, we start with a simple toy model to explain what we want to achieve.

\subsection{General Idea}
\label{sec:tftof:idea}
Imagine a minimalistic experimental setup with a source at position $z_\up{S}=0$ that emits electrons or ions of different energies parallel to the z-axis and a detector at $z_\up{D}$ that can only count the charged particles and record their arrival time, but cannot measure their energy.
For each electron or ion the start time is unknown since the source is continuously emitting them.
We can set up a drift tube of length $L$ between detector and source at position $z_1$ and apply a time-dependent electric potential $U(t)$ to it.
\begin{figure}[hbt]
	\centering
	\includegraphics[width=\linewidth]{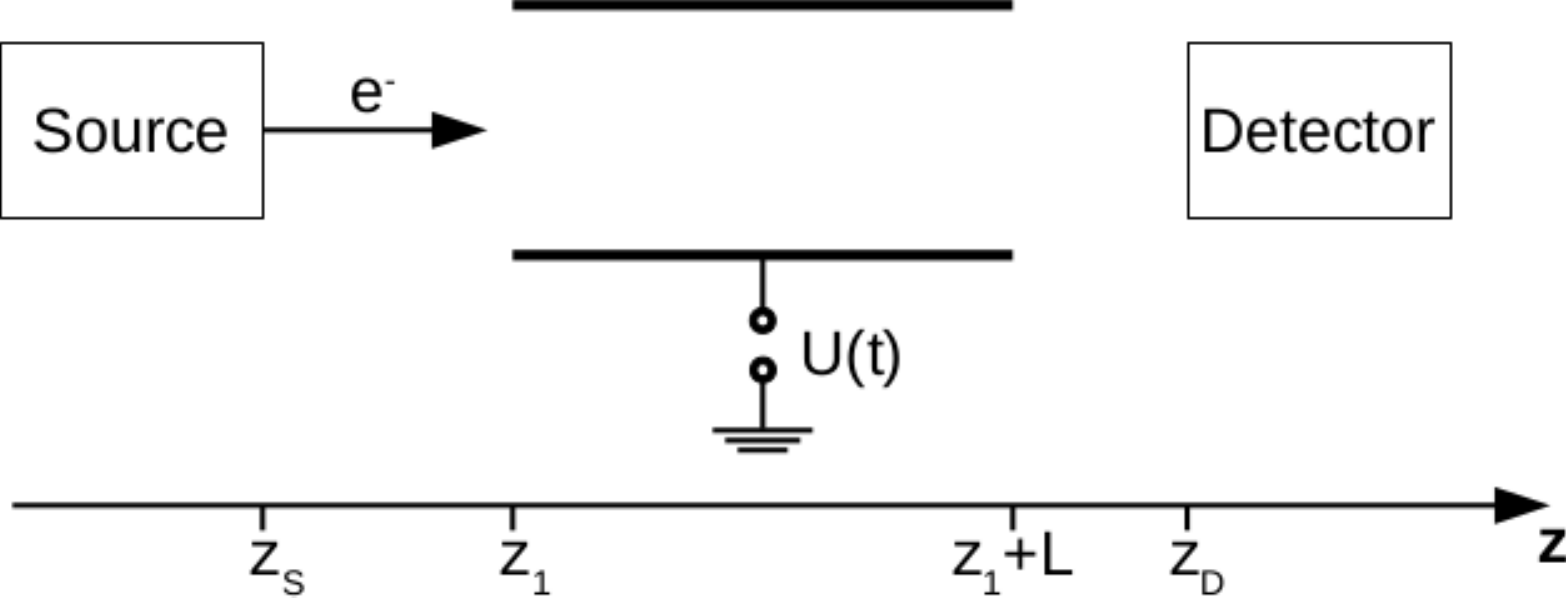}
	\caption{Schematic illustration of the toy model setup.}
	\label{fig:toy-setup}
\end{figure}
The setup is depicted in figure~\ref{fig:toy-setup}.
For simplicity we assume the length of the drift tube to be much larger than its radius, such that the electric field is only non-zero in a negligible short range at the entrance $z_1$ and exit $z_1+L$ of the drift tube.
The electric potential for the system now looks like
\begin{equation}
	\label{eq:usimple}
U(z,t) = \begin{cases}
    0 & \text{, } z \leq z_1 \\
    U(t) &\text{, } z_1 < z < z_1+L \\
    0 & \text{, } z_1+L \leq z
\end{cases} \ .
\end{equation}
\noindent The flight time $\tau$ of electrons or ions with initial kinetic energy $E_0$ and total relativistic energy $E_\up{rel}=E_\up{0}+mc^2$ with $m=m_\up{e} \text{ or } m_\up{ion}$ over a distance $l$ is given by
\begin{equation}
\label{eq:flight_time}
\tau = \frac{l}{v}=\frac{l}{\beta c} = \frac{l E_\up{rel}}{pc^2}=\frac{l E_\up{rel}}{c\sqrt{E_\up{rel}^2-m^2c^4}}\ ,
\end{equation}
where $v$ and $p$ are the velocity and momentum of the particle, respectively.
For the individual sections of the model defined by eq.~\eqref{eq:usimple}, the corresponding flight times $\tau_i$ of electrons or ions starting at time $t_0'$ can be expressed as
\begin{align}
	\label{eq:tausimple}
	\tau_1 = &  \frac{z_1-z_\up{S}}{c}\frac{E_\up{rel}}{\sqrt{E_\up{rel}^2-m^2c^4}} \\
	\tau_2 = &  \frac{L}{c}\frac{E_\up{rel}-qU(t_0'+\tau_1)}{\sqrt{(E_\up{rel}-qU(t_0'+\tau_1))^2-m^2c^4}} \\
	\tau_3 = &  \frac{z_\up{D}-(z_1+L)}{c}\times \\
	 & \frac{E_\up{rel}-q[U(t_0'+\tau_1)-U(t_0'+\tau_1+\tau_2)]}{\sqrt{(E_\up{rel}-q[U(t_0'+\tau_1)-U(t_0'+\tau_1+\tau_2)])^2-m^2c^4}}\ , \nonumber
\end{align}
\noindent where the TOF inside the drift tube $\tau_2$ depends on the applied potential $U$ at the time $ t_0'+\tau_1$ the electron (ion) enters it.
The arrival time $t_\up{arr}$ is the sum of all flight times and by choosing $t_0 \equiv t_0'+\tau_1$, it reads as
\begin{equation}
    \label{eq:arrivalsimple}
    t_\up{arr}(t_0, E_0) = t_0 +\tau_2+\tau_3 \ .
\end{equation}
\noindent In principle $\tau_3$ depends on the change of the electric potential $U$ during the time $\tau_2$. However, if the length of the drift tube is chosen sufficiently large compared to the distance between the tube and the detector ($L\gg z_D-(z_1+L)$), $\tau_3$ can be neglected. This is further justified by the usually high electron (ion) velocities outside of the drift tube\footnote{Assuming the proposed experimental implementation in KATRIN discussed in sec. \ref{sec:katrin}, electrons take $\tau_2\approx \SI{10}{\micro\second} $ to pass the $L=\SI{24}{\meter}$ tfTOF section, but only $\tau_3 \approx \SI{30}{\nano\second}$ to travel the $z_D-(z_1+L)=\SI{2}{\meter}$ distance from the exit of the drift tube to the detector.}.
For the remaining part of this section, $\tau_3$ is therefore neglected, while the simulations carried out in section \ref{sec:katrin} take $\tau_3$ into account.
Equation~\eqref{eq:arrivalsimple} then reduces to
\begin{equation}
    \label{eq:arrivalsimpler}
    t_\up{arr}=t_0+\frac{L}{c}\frac{E_\up{rel}-qU(t_0)}{\sqrt{(E_\up{rel}-qU(t_0))^2-m^2c^4}} \\
\end{equation}

\noindent which can be rearranged for $U(t_0)$:
\begin{equation}
    \label{eq:accvolt_rel}
    U(t_0) = -\frac{1}{q}\left(\frac{mc^2(t_\up{arr}-t_0)}{\sqrt{(t_\up{arr}-t_0)^2-L^2/c^2}}-(E_0+mc^2)\right)\ ,
\end{equation}

\noindent or in the non-relativistic limit:
\begin{equation}
    \label{eq:accvolt}
    U(t_0) = -\frac{1}{q}\left(\frac{mL^2}{2(t_\up{arr}-t_0)^2}-E_\up{0}\right)\ .
\end{equation}

\begin{figure}[hbt!]
	\centering
	\includegraphics[width=\linewidth]{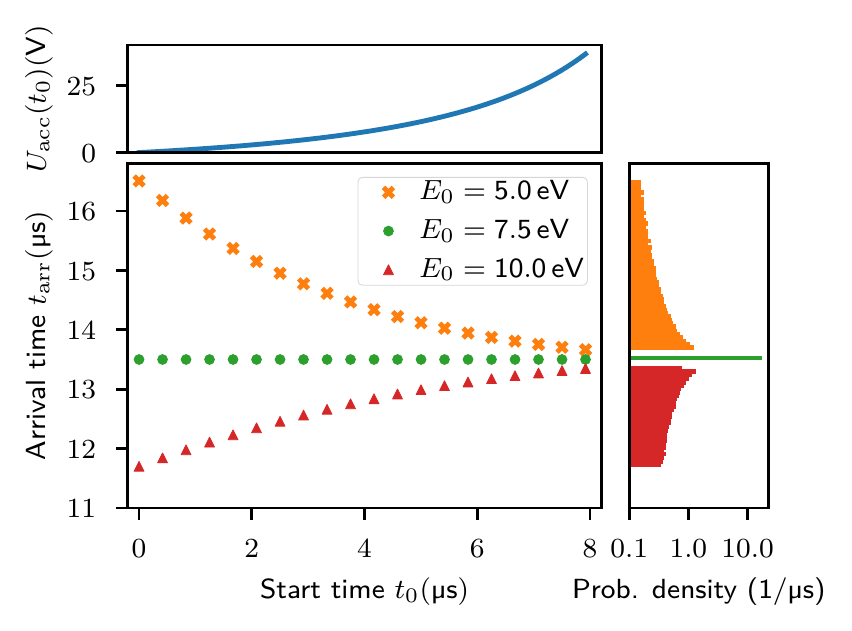}
	\caption{Start time dependent arrival time for three different electron energies in a toy model setup with a $L=\SI{22}{\meter}$ long drift tube and an applied accelerating voltage (top) according to eq.~\eqref{eq:accvolt} for $E_\up{foc}=\SI{7.5}{\electronvolt}$ and $t_\up{foc}=\SI{13.5}{\micro\second}$.
	The histogram (right) shows the probability density for an electron of given energy to arrive at a certain time.
	For the focused energy this is an infinitely sharp distribution, while for other energies it is smeared out more.}
	\label{fig:tarr-spec-toy}
\end{figure}

If we now fix a kinetic energy $E_\up{foc}\mathrel{\mathop:}=E_\up{0}$ and an arrival time $t_\up{foc}\mathrel{\mathop:}=t_\up{arr}$ in eq.~\eqref{eq:accvolt_rel} and apply this potential to our drift tube, all electrons (ions) of the chosen energy will arrive at time $t_\up{foc}$, independent of their start time $t_0$.
This focusing of particles with common energy but different start times is what we call time-focusing, and it allows for a distinction of different energies by their arrival times as illustrated in figure~\ref{fig:tarr-spec-toy}.

For times $t_0\rightarrow t_\up{foc}$ eq.~\eqref{eq:accvolt} diverges, which can be prevented if the applied potential is periodized with period $T<t_\up{foc}$.
The timing parameters $T$ and $t_\up{foc}$ determine the minimum and maximum acceleration voltage and influence how strongly the arrival times for different energies are separated.
In practice, it is desirable to have a low ramping amplitude due to the non-negligible capacitance of the drift tube.
That can be achieved by a sufficiently small ramping period and a high focal time $t_\up{foc}$.
However, if they are set too small or too large respectively, it becomes increasingly difficult to distinguish the energies due to overlapping arrival times from particles starting in different periods.

For a sinusoidal voltage $U(t_0)=U_0\sin{(2\pi f t_0)}$, the reactive power $P$ is given by
\begin{equation}
\label{eq:power}
P=2\pi\cdot C\cdot U_0^2\cdot f \ ,
\end{equation}
\noindent where $C$ is the capacitance of the drift tube.
A reactive power also leads to real power dissipation in the cables connecting the voltage supply with the tube.
Therefore, the output power of the power supply needs to be large enough in order to provide a sinusoidal voltage with frequency $f$ and amplitude $U_0$.
In case of a non-sinusoidal voltage like in eq.~\eqref{eq:accvolt}, this effectively defines the maximum amplitude for each frequency component in Fourier space, which can lead to a smearing of the function if the output power is not sufficient.

\subsection{Time-Focusing Time-of-Flight in a MAC-E-Filter}
\label{sec:tftof:mac-e}
The toy model discussed in the previous section represents an idealized, somewhat oversimplified setup to demonstrate the general idea of the tfTOF method.
It allows to identify some key points for the method to work, which are:
\begin{itemize}
\item low electron (ion) velocities,
\item collimated momenta,
\item sufficiently long drift tube and
\item nearly vanishing electric field in the delay line.
\end{itemize}
Those requirements are approximately met inside a MAC-E-filter, whose retarding potential slows down electrons, while the momenta are adiabatically collimated due to the magnetic field configuration (as will be shown below in eq.~\eqref{eq:p_perp}).
With a length of roughly \SI{24}{\meter} the state-of-the-art MAC-E spectrometer of the KATRIN experiment also demonstrates that the assumed length of the drift tube used in the toy model can be realized.

In order to obtain the TOF of an electron (ion) as a function of the initial kinetic energy $E_0$ and polar angle $\theta_0$ in a MAC-E-filter type drift tube as delay line, one has to integrate the reciprocal velocity over the center of motion.
For the simple case of an electron (ion) created on the rotational symmetry axis $z$ of the tube, this reads \cite{Steinbrink2013}:

\begin{equation}
	\label{eq:tof}
	\mathcal T (E_0, \theta_0) = \int \up d z \ \frac{1}{v_\parallel} 
	= \int\limits_{z_\up{0}}^{z_\up{arr}} \up d z \ \frac{E_\up{kin}(z) + mc^2}{p_\parallel(z, \theta_0) \cdot c^2}\ ,
\end{equation}

\noindent where $z_\up{0}$ and $z_\up{arr}$ are the positions on the beam axis between which the TOF is measured, $E_\up{kin}(z)$ is the kinetic energy at longitudinal position $z$, which is given by the initial kinetic energy $E_0$ plus the work done by the electric field, and $p_\parallel(z, \theta_0)$ is the parallel momentum at position $z$:
\begin{equation}
	\label{eq:p_parallel}
	p_\parallel(z, \theta_0)c = \sqrt{E_\up{rel}^2(z)-p_\perp^2(z,\theta_0)c^2-m^2 c^4}\ .
\end{equation}

\noindent Here, $E_\up{rel}=E_\up{kin}(z)+m c^2$ is the relativistic energy of the particle at position $z$ and $p_\perp^2(z)$ is the transverse momentum which itself can be written as function of polar angle, momentum and magnetic field in the source as well as $B(z)$ in the case of adiabatic motion:
\begin{equation}
\label{eq:p_perp}
p_\perp^2(z,\theta_0) = p_0^2\sin^2(\theta_0)\frac{B(z)}{B(z_0)}\ .
\end{equation}

For the non-focusing TOF method \cite{Steinbrink2013}, the kinetic energy can simply be expressed by

\begin{equation}
    \label{eq:ekinstandard}
    E_\up{kin}(z) = E_0 - q \Delta U(z) \ ,
\end{equation}

\noindent where $E_0$ denotes the initial kinetic energy at $z=z_0$ and $\Delta U(z)$ the potential difference between $z_0$ and $z$.
However, with tfTOF, the electric potential becomes time-dependent:

\begin{equation}
	\label{eq:timedep}
\Delta U(z) \rightarrow \Delta U(z, t)\ .
\end{equation}

\noindent Therefore, eq.~\eqref{eq:ekinstandard} holds no longer true, since the potential also changes over the flight path of the electron (ion) due to its time-dependency.
The kinetic energy now can be expressed by

\begin{equation}
    \label{eq:ekintrue}
    E_\up{kin}(z) = E_0 - q \int_{z_0}^z \up d z' \ \frac{\partial U(z', t(z'))}{\partial z'} \ .
\end{equation}

\noindent Because of the time-dependency of the electric potential, the TOF now also depends on the start time $t_0=t(z_0)$:

\begin{equation}
	\label{eq:toftimedep}
\mathcal T (E_0, \theta_0) \rightarrow \mathcal T (E_0, \theta_0, t_0) \ .
\end{equation}

\noindent The quantity measured with the tfTOF method is the arrival time $t_\up {arr}$ at the detector which is given by the start time $t_0$ plus the TOF:
\begin{equation}
	\label{eq:toa}
t_\up{arr} (t_0, E_0, \theta_0) = t_0 + \mathcal T (E_0, \theta_0, t_0)\ .
\end{equation}

In general, eq.~\eqref{eq:tof} is not solvable analytically, since the potential $\Delta U(z)$ for given time $t$ and the magnetic field $B(z)$ are complicated functions of the experimental geometry.
Moreover, for a time-dependent electric potential, eq.~\eqref{eq:tof} becomes a complicated integral equation.

The achievable energy resolution of TOF methods is ultimately limited by the time resolution of the detector system, which for the KATRIN focal plane detector is about \SI{50}{\nano\second} \cite{Amsbaugh2015}.
Since the TOF $\mathcal{T}$ scales with $\frac{1}{\sqrt{E_\up{kin}}}$ according to eq.~\eqref{eq:tof}, a given energy difference $\Delta E$ yields a larger flight time difference $\Delta \mathcal T$ for smaller absolute kinetic energies $E_\up{kin}$.
Electrons or ions should therefore have small energies on the order of eV to be able to get a sub-eV energy resolution from TOF methods.
In addition, the flight time scales linearly with the traveled distance $z$, which makes longer delay lines favorable.

Because eq.~\eqref{eq:tof} depends not only on the kinetic energy $E_\up{kin}$, but also on the starting angle $\theta_0$, the time-of-flight can be exactly the same for two electrons of different energies and angles.
To break the energy degeneracy, the momenta should be collimated as much as possible between $z_0$ and $z_\up{arr}$, which is the case inside a MAC-E-filter.

It should be mentioned that a MAC-E-filter in tfTOF mode cannot be used on its own in most circumstances because the energy resolution of the integral measurement would worsen by the amplitude of the acceleration voltage.
Instead, a tandem of conventional MAC-E-filter with a static retarding voltage, followed by one with a ramped potential is able to combine the benefits of both systems.


\section{Application of the tfTOF Method to the KATRIN Experiment}
\label{sec:katrin}
The implementation of a tfTOF mode into KATRIN has several advantages.
Given a feasible means of introducing a ramping section into the setup, we can obtain the electron arrival time spectra without impairing the integrating standard MAC-E-filter mode.
In addition to the count-rate, a full arrival time spectrum is then gained for each retarding potential, allowing to study spectral features such as the neutrino mass cut-off or a sterile neutrino contribution in more detail.
This surplus of information is expected to increase the statistical sensitivity and may possibly reduce systematic uncertainties as well.

\subsection{Experimental Implementation of a tfTOF section at the KATRIN Experiment}

\begin{figure*}[hbt]
	\centering
	\includegraphics[width=\linewidth]{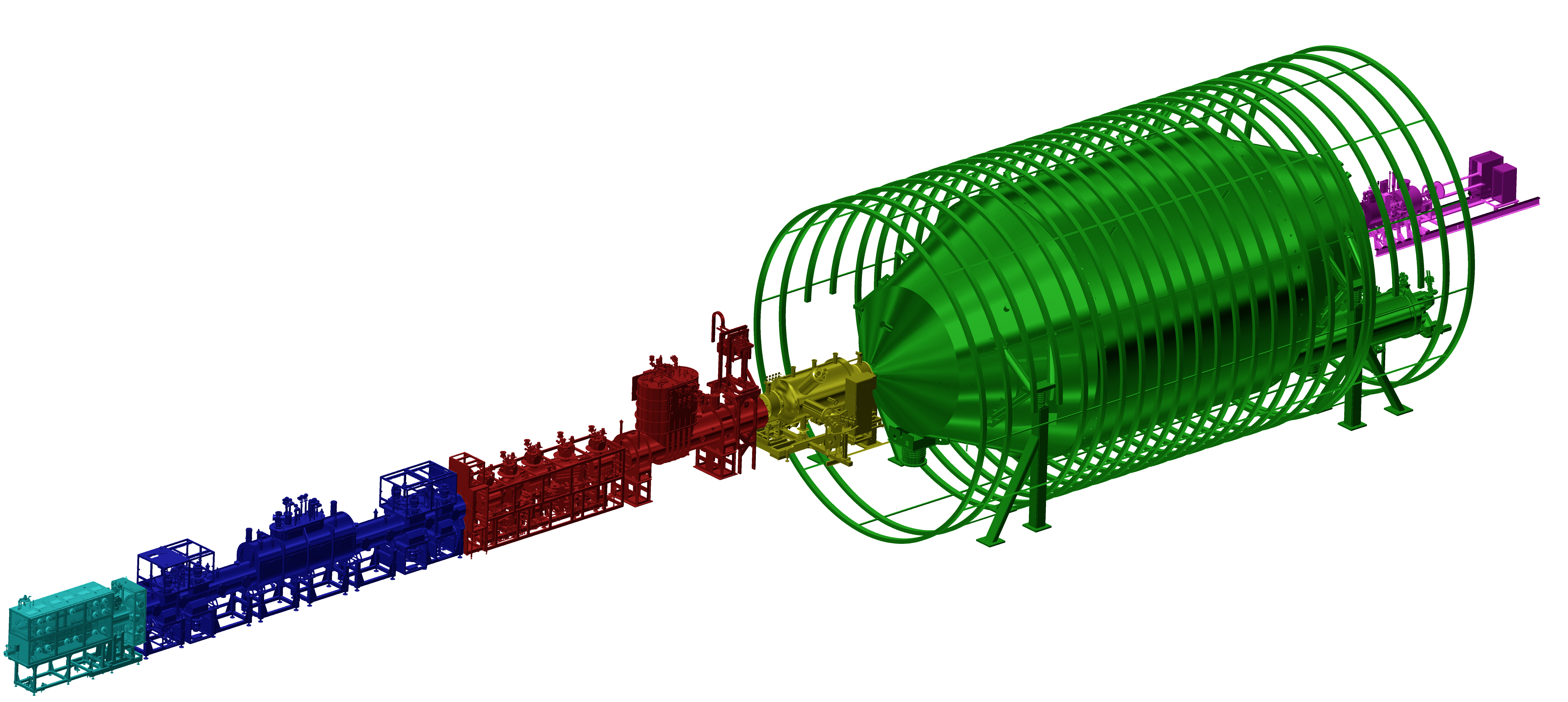}
	\caption{Overview over the \SI{70}{m} long KATRIN beam-line.
	From left to right the different components are: Calibration and Monitoring Section (cyan), Windowless Gaseous Tritium Source (blue), Differential Pumping and Cryogenic Pumping Sections (red), Pre-Spectrometer (yellow), Main Spectrometer (green), Detector Section (purple) \cite{Aker2019}. The additional tfTOF hardware section (not shown) is assumed to be placed between the main spectrometer and the detector section and to be of the same size as the main spectrometer, increasing the beamline length to \SI{94}{m}.}
	\label{fig:beamline}
\end{figure*}

KATRIN has been designed to measure the endpoint region of the $\beta$-decay of tritium.
We use a detailed modeling of the experiment in general and of the $\beta$-spectrum and MAC-E filter in particular to obtain realistic results.
The differential tritium decay rate for a kinetic energy $E$ of the electron can be written as \cite{Otten2008}
\begin{multline}
\label{eq:beta_spec}
\frac{\up{d}\dot{N}}{\up{d}E} (E) = N \frac{G_\up{F}^2}{2\pi^3\hbar^7 c^5}\cos^2(\theta_\up{C})\left|M\right|^2 F(E,Z')p(E+m_\up{e} c^2) \\
\times \sum_{i}P_i(E_0-V_i-E)\sqrt{(E_0-V_i-E)^2-m^2(\nu_e)c^4}\ ,
\end{multline}
\noindent with the total number of tritium atoms in the source $N$, the Fermi constant $G_\up{F}$, the Cabbibo angle $\theta_\up{C}$, the nuclear matrix element $M$, the Fermi function $F(E,Z')$ which takes into account the charge of the daughter nucleus $Z'$, the momentum of the electron $p$, the probability $P_i$ for the molecule to decay into an electronically or rotational-vibrationally excited state with excitation energy $V_i$ \cite{Doss2006} and the endpoint energy $E_0$, which would be the maximum kinetic energy of the electron in case of $m(\nu_e)=0$.
The parameter of interest for KATRIN is the average electron neutrino mass 
\begin{equation}
\label{eq:neutrino_mass}
m(\nu_e) \mathrel{\mathop:}= \sqrt{\sum_{i} \left|U_{\up{e}i}\right|^2m_i^2}\ ,
\end{equation}
which is the incoherent sum over the three neutrino mass eigenstates $m_i$, taking into account their individual contributions $U_{\up{e}i}$ to the electron neutrino mass.

In the KATRIN experiment, shown in figure \ref{fig:beamline}, $\beta$-electrons from the tritium decay inside the windowless gaseous tritium source (WGTS \cite{KATRIN2005}) are magnetically guided through the pumping section towards the MAC-E spectrometers, while the tritium is pumped out first differentially and then cryogenically with a total retention factor of $>10^{14}$ \cite{Aker2019}.
The MAC-E-filter type main spectrometer analyzes the energy of the electrons, before they are detected at the focal plane detector (FPD).
The $\beta$-electron rate $\dot{N}_\up{S}$ at a given $qU$ value is hereby given by a convolution of eq.~\eqref{eq:beta_spec} with the response function $R(E,qU)$ of the experiment
\begin{equation}
\label{eq:expected_rate}
\dot{N}_\up{S}(qU)= \frac{\Delta \Omega}{4\pi}\int_{0}^{E_0}\frac{\up{d}\dot{N}}{\up{d}E}(E)R(E,qU)\up{d}E\ .
\end{equation}
\noindent Here, $\frac{\Delta \Omega}{4\pi}=\frac{1}{2}(1-\cos(\theta_{max}))$ is the accepted solid angle with a value of $\approx 0.18$ in the case of KATRIN.
The response function itself is a convolution of the MAC-E-filter transmission function, see eq.~\eqref{eq:transmission}, with the energy loss function \cite{Aseev2000,Abdurashitov2017}, which arises from the fact that electrons inside the source can lose energy by scattering with other tritium molecules.

In the following section, the prospects of an additional dedicated tfTOF hardware section with the same geometry and electromagnetic properties as the existing KATRIN main spectrometer are investigated.
It is placed between the main spectrometer and the detector section so that it only has to analyze a small energy region $qU_\up{MS}\leq E \leq E_0$ between the retarding potential of the main spectrometer $U_\up{MS}$ and the endpoint of the tritium $\beta$-spectrum $E_0$.
This is intended to give a reasonable estimate on what the benefit of such a section would be, while at the same time it still leaves room for optimizations regarding the cost-benefit ratio.

\subsection{tfTOF Spectrum and Neutrino Mass Sensitivity}
\label{sec:katrin:sensitivity}
In order to investigate the benefits of tfTOF on KATRIN's sensitivity, the generation of neutrino mass dependent time-of-arrival (TOA) spectra of the tritium $\beta$-decay electrons is necessary.
Because of the complex geometry and time dependence of the electric potential and field inside the spectrometer, a Monte Carlo approach was chosen over an analytical one.
The analysis has been carried out in a modular fashion to allow for more flexibility by separating the computationally expensive particle tracking simulations from the less time consuming spectrum generation and fitting.
The different parts are described in the following subsections.

\subsubsection{Particle Tracking Simulation}
\label{sec:katrin:sensitvity:sim}
The first step towards a $\beta$-electron TOA spectrum is the particle tracking inside the tfTOF section.
Here, we use the \textsc{Kassiopeia} \cite{Furse2017} framework, which has been developed within and is maintained by the KATRIN collaboration.
An axial symmetric model of the KATRIN main spectrometer with adapted potential settings of the inner electrodes is simulated.
For this study, a range of the last \SI{25}{\electronvolt} below the tritium endpoint $E_0=\SI{18575}{\electronvolt}$ is selected.
The ramping parameters of the potential are estimated by the simple toy model for a length of $L=\SI{22}{m}$, an arrival time of $t_\up{arr}=\SI{13.5}{\micro\second}$ for purely forward emitted electrons of surplus energy $E_\up{foc}=\SI{7.5}{\electronvolt}$ and a period of $T=\SI{8}{\micro\second}$, resulting in a maximum value for the acceleration voltage of 
\begin{equation}
\lim\limits_{t \rightarrow T}{U_\up{acc}(t)}=\SI{38}{\volt}\ ,
\end{equation}
according to eq.~\eqref{eq:accvolt}.

The electric potentials in the tfTOF section are calculated once with \textsc{Kassiopeia}'s field calculation module \emph{KEMField} for the minimal $U_\up{min}=U_\up{MS}$ and maximal $U_\up{max}=U_\up{MS}+\lim\limits_{t\rightarrow T} U_\up{acc}(t)$ electrode settings.
During the tracking the potential at time $t$ is given by superposition of the static solutions according to
\begin{equation}
\label{eq:superposition}
U(t)=(1-a(t))\cdot U_\up{min}+a(t)\cdot U_\up{max}\,,
\end{equation}
where $a(t)$ is a time dependent modulation, which in our case is:
\begin{equation}
\label{eq:modulation}
a(t) = \frac{(t_\up{foc}-T)^2}{t_\up{foc}^2-(t_\up{foc}-T)^2}\cdot\left(\frac{t_\up{foc}^2}{(t_\up{foc}-t \bmod T)^2}-1\right)\ .
\end{equation}
This modulation has the same time dependence as eq.~\eqref{eq:accvolt} an fulfills the conditions $a(0)=0$ and $\lim\limits_{t\rightarrow T}{a(t)}=1$ imposed by the superposition approach in eq.~\eqref{eq:superposition}.

For the actual tracking, electrons are started on the symmetry axis in front of the spectrometer with a fixed energy, for 360 different start times $t_0$ equally distributed throughout one period $T=\SI{8}{\micro\second}$ of the ramping voltage.
At each sample time electrons are started with 101 different angles towards the magnetic field $\theta_0$, according to an isotropic distribution in the source up to $\theta_\up{max}$.
This is repeated four times, resulting in $\approx 1.5\cdot10^5$ electrons with an azimuthal starting angle randomly chosen between \SI{0}{\degree} and \SI{360}{\degree}.
This procedure is carried out for 168 different energies which are distributed such that for each interval the same number of electrons from the tritium $\beta$-decay is expected, but the maximal distance between two energies is not larger than \SI{200}{\milli\electronvolt}.
\begin{figure}[hbt]
  \centering
  \includegraphics[width=\linewidth]{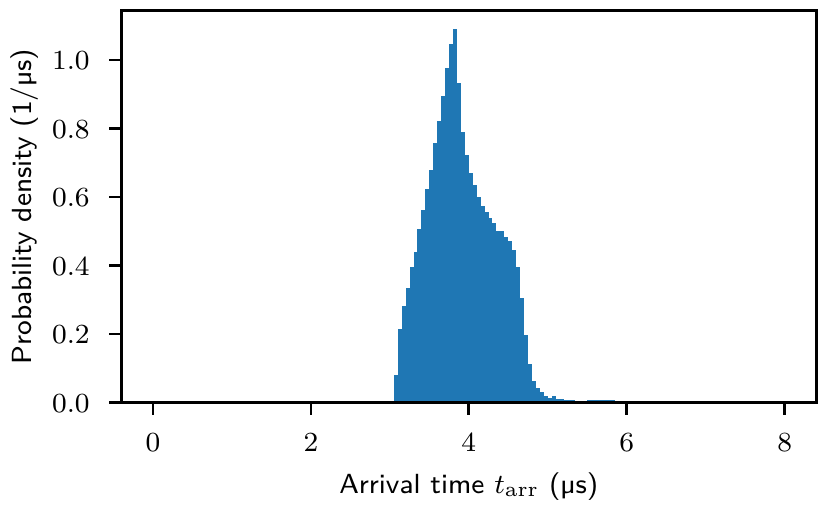} 
  \caption{Time-of-arrival distribution for isotropically emitted (up to $\theta_\up{max}$, see eq.~\eqref{eq:max_start_angle}), monoenergetic electrons with starting energies \SI{3}{\electronvolt} above the transmission edge $qU_\up{MS}$ of the regular main spectrometer and with uniform start time distribution.
  The data is binned with a \SI{50}{\nano\second} width to account for the time resolution of the detector.}
  \label{fig:response}
\end{figure}

Finally, the arrival time for each event is taken modulo the period and a TOA histogram with the probability density is created for each energy.
The resulting histograms are the response of the system to electrons with a given energy, taking into account the angular and temporal distributions.
An example response for electrons with \SI{3}{\electronvolt} surplus energy can be seen in figure~\ref{fig:response}.

\subsubsection{Time-of-Arrival $\beta$-Spectrum}
\label{sec:katrin:sensitivity:spectrum}
\begin{figure}[hbt]
  \centering
  \includegraphics[width=\linewidth]{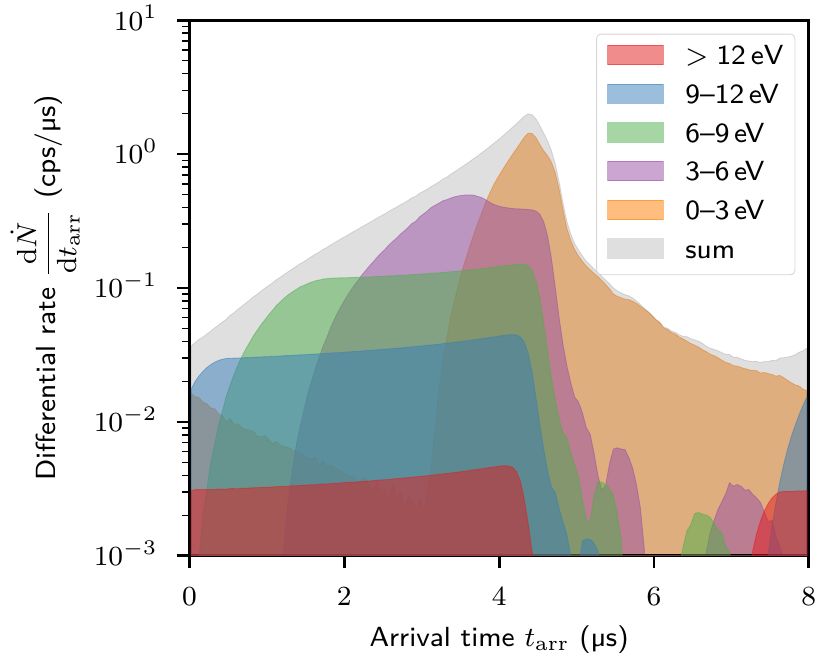} 
  \caption{Total time-of-arrival spectrum of $\beta$-electrons for $m(\nu_e)=\SI{0}{\electronvolt}$ and $E_0=\SI{18575}{\electronvolt}$ at a single main spectrometer retarding energy $E_0-qU_\up{MS}=\SI{16.3}{\electronvolt}$ below the endpoint (gray) with several surplus energy ranges $E-qU_\up{MS}$ in color code to visualize that each electron energy has a distinct impact on the overall spectrum. The electrons start uniformly distributed in time and with an isotropic angular distribution up to $\theta_\up{max}$.}
  \label{fig:tarr-spectrum}
\end{figure}

\begin{figure}[hbt]
  \centering
  \includegraphics[width=\linewidth]{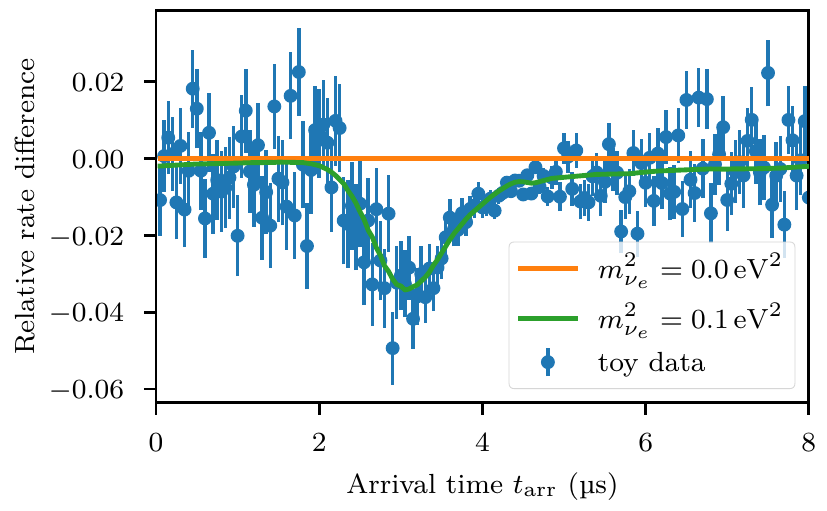} 
  \caption{Influence of a non-zero neutrino mass on the spectrum in terms of relative rate difference for a single retarding potential $\SI{6.3}{\electronvolt}$ below the endpoint.
  The blue points represent toy data generated for three years of measurement time and an isochronous background rate of \SI{10}{\milli\hertz} for $m^2(\nu_e) = 0.1$~ eV$^2$, whereas the green and orange lines represent the underlying model for two squared neutrino mass values. Since the endpoint $E_0$ is fixed to illustrate the dependence on $m^2(\nu_e)$, the statistical sensitivity cannot be derived from this simple plot.}
  \label{fig:tarr-neutrino-mass-influence}
\end{figure}

With the obtained TOA spectra for each energy we can now build a neutrino mass dependent TOA model.
We do this by sampling electron energies according to the expected distribution from the tritium decay in eq.~\eqref{eq:beta_spec} and combining the TOA spectra for the two nearest simulated energies through linear interpolation.
The result can be seen in figure~\ref{fig:tarr-spectrum} and the influence of the neutrino mass on the spectrum is illustrated in figure~\ref{fig:tarr-neutrino-mass-influence}.

To enable the investigation of the sensitivity to the neutrino mass for a real experiment later, we will need to fit simulated TOA spectra w.r.t. the squared neutrino mass $m^2(\nu_e)$ and endpoint energy $E_0$.
Therefore, this approach is used for combinations of different neutrino masses and endpoints, resulting in a grid of TOA spectra that can be used as models.
Such a grid can be generated for different retarding potentials of the main spectrometer and tfTOF section as well.
Choosing several different retarding potentials close to the endpoint is likely to reduce systematics and has also shown better statistical sensitivity in the ideal TOF case (cf. \cite[chapter~4]{Steinbrink2013}).
For this analysis we decided to take four different potentials at \SI{6.3}{\electronvolt}, \SI{11.3}{\electronvolt}, \SI{16.3}{\electronvolt}, \SI{26.3}{\electronvolt} below the assumed endpoint of $E_0=\SI{18575}{\electronvolt}$.

\subsubsection{Sensitivity Analysis}
\label{sec:katrin:sensitivity:analysis}
In order to investigate the benefits of tfTOF on the neutrino mass sensitivity, we generate a toy data set with a given $E_0$ and $m^2(\nu_e)$ chosen from the grid, scaling it with the measurement time and sampling the obtained counts in each bin according to a Poisson distribution.
To perform a $\chi^2$ optimization on all four relevant parameters -- neutrino mass squared $m^2(\nu_e)$, endpoint energy $E_0$, background rate and signal amplitude -- for this Monte Carlo data set, we proceed in two steps.
First, for each TOA spectrum on the grid a $\chi^2$ optimization of background and signal amplitude parameters is performed. Second, the minimum of the $\chi^2$ manifold in the parameter space  $m^2(\nu_e)$ versus $E_0$ is noted down. In this way we obtain the global $\chi^2$ minimum in the four parameter space with discretized $m^2$ and $E_0$. The process of toy data generation and two step $\chi^2$ optimization is repeated for 1000 times to assess the sensitivity by an ensemble test.
The statistical sensitivity for the neutrino mass can then be estimated by the standard deviation of the $m^2(\nu_e)$ values of the grid points showing the best $\chi^2$.

\begin{table}[hbt]
    \centering
	\caption{Comparison of statistical neutrino mass sensitivities for different TOF methods with the standard integral measurement modes under KATRIN design conditions \cite{KATRIN2005} (1$^\text{st}$ row) and at the higher background rate of KATRIN's first science run (2$^\text{nd}$ row) scaled with the simple scaling laws given in \cite{KATRIN2005} (footnote 34) and in \cite{Otten2008} (equation 48). For the full TOF method (3$^\text{rd}$ row) using a hypothetical electron tagger discussed in \cite{Steinbrink2013}, a background of \SI{0}{\milli\hertz} is assumed because it can be separated from signal electrons by coincidence methods.
	The two tfTOF cases investigated here either use an isochronous background rate of \SI{10}{\milli\hertz} (4$^\text{th}$ row), relating to the KATRIN design conditions, or an isochrounous background rate of \SI{42}{\milli\hertz} combined with a \SI{261}{\milli\hertz} time-focused contribution from the main spectrometer (5$^\text{th}$ row).}
		\label{tab:results}
	\begin{tabular}{c c c}
		\hline
		Method &  \vtop{\hbox{\strut Background}\hbox{\strut (\SI{\milli\hertz})}}& \vtop{\hbox{\strut statistical error}\hbox{\strut $\sigma_\up{stat}\left(m^2(\nu_e)\right  )$ (\SI{\electronvolt})}}  \\
		\hline
		standard  & 10 & 0.018 \cite{KATRIN2005}\\
		standard  & 293\footnotemark & 0.032 - 0.035 \cite{KATRIN2005,Otten2008} \\
		TOF  & 0 & 0.0032 \cite{Steinbrink2013}\\
		tfTOF  & 10 & 0.009 [this work]\\
		tfTOF  & 42 + 261 & 0.015 [this work]\\
		\hline
	\end{tabular}
\end{table}
\footnotetext{This was the background rate during KATRIN's first science run. Meanwhile the background rate could be decreased to around 100~mcps by baking the spectrometer and using a special electromagnetic setting. The inner electrode system of the main spectrometer can be used to shift the analyzing plane (''SAP'') towards the detector, decreasing the volume dependent "Rydberg electron" background \cite{PhDDyba2018}.}
Table \ref{tab:results} shows the results obtained by our simulations for a MC truth of $m^2(\nu_e)=0~\up{eV}^2$ and $E_0=\SI{18575}{\electronvolt}$. 
Using the method described in the last paragraph for the tfTOF scenario, we find a statistical neutrino mass sensitivity of $\sigma^\up{tfTOF}_\up{stat}(m^2(\nu_e))=0.009~\up{eV^2}$ after three years of live time, when we assume an isochronous background of $\SI{10}{\milli\hertz}$.
This value is chosen to compare the new tfTOF method with the  KATRIN design sensitivity $\sigma^\up{std}_\up{stat}(m^2(\nu_e))=$ 
\SI{0.018}{\electronvolt\squared} \cite{KATRIN2005}, which is given for the same background rate.
Unfortunately, KATRIN has not achieved this low background yet.
Instead, the rate during its first science run has been $R_\up{bg}=\SI{293}{\mcps}$ \cite{Aker2019} and about \SI{200}{\mcps} in its second science run after out-baking the spectrometer.
Therefore, a much more interesting comparison is the one between the standard operation mode with a background rate of \SI{293}{mcps} and the tfTOF mode with a high background rate of \SI{261}{\mcps} from the main spectrometer, \SI{32}{\mcps} from radioactivity nearby the detector or other sub-dominant background sources and an additional background rate of \SI{10}{mcps} from the new tfTOF section (for the latter background rate we assume that we can build the tfTOF section under radon-free atmosphere avoiding the ``Rydberg electron'' background, see below).

While the \SI{32}{\mcps} background and the additional background electrons from the tfTOF section are assumed to be uniformly distributed across all arrival times, the main spectrometer background contributes a non-homogeneous component to the time-of-arrival spectrum.
This comes from the fact that this part of the background also undergoes the time-focusing in the tfTOF section.
Thus, the energy and angular distributions of the spectrometer background rate play a crucial role in how it impacts the overall spectrum.

By several investigations, e.g. artificial injection of $^{219}$Rn and $^{220}$Rn into KATRIN's main spectrometer, it was concluded that the main component of the elevated background rate of \SI{293}{\mcps} or \SI{200}{\mcps} in the first and second science run, respectively, originates from the so-called ``Rydberg electron'' background  \cite{Fraenkle2017,PhDTrost2019}: During the installation of the KATRIN main spectrometer the $\alpha$-decay of $^{214}$Po, a decay daughter of $^{222}$Rn in the air and in materials, implanted the long-lived isotope $^{210}$Pb ($t_{1/2} = \SI{22.3}{yr}$) into the vessel walls, resulting in a measured $^{210}$Pb surface activity of about 1~mcps/m$^2$. 
The $\alpha$-decay of the $^{210}$Pb decay daughter $^{210}$Po has a non-zero probability to eject neutral atoms in a highly excited state (``Rydberg'' atoms) into the vacuum passing all electric and magnetic barriers on their way into the relevant magnetic flux tube of the spectrometer. By absorption of an infrared photon from the thermal radiation of the vessel wall at room temperature, eventually the Rydberg atoms get ionized releasing very low energy electrons. 
\begin{figure}[hbt]
  \centering
  \includegraphics[width=\linewidth]{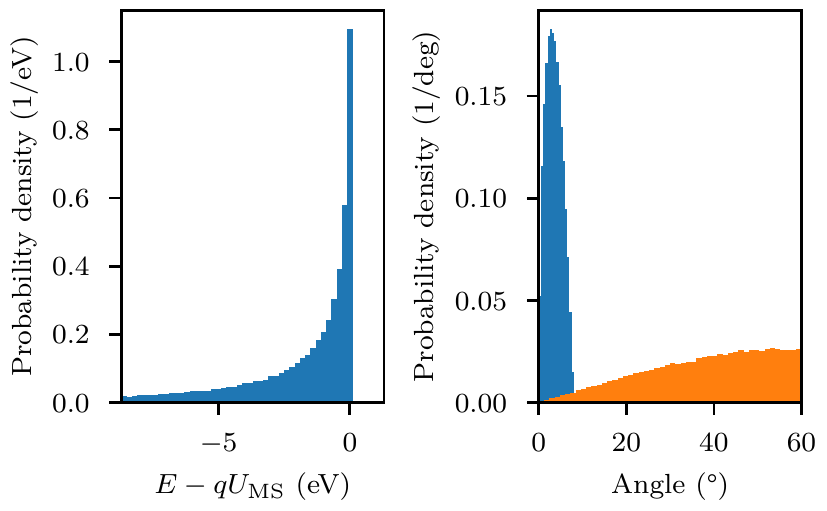} 
  \caption{Energy (left) and angular (right) distribution of the ``Rydberg background'' electrons (blue) at the entrance of the tfTOF section. The right plot also shows the angular distribution of beta electrons at the same position (orange).}
  \label{fig:toa_energy_angle}
\end{figure}
These  ``Rydberg background'' electrons have very little initial energies, $\cal{O}$$(k_BT = \SI{25}{meV}$) and -- depending on their start position -- get electrostatically accelerated to either end of the main spectrometer. Therefore they have a distinguished energy distribution around the retarding energy $qU_\up{MS}$ and possess a sharp angular distribution around zero degree (see fig. \ref{fig:toa_energy_angle}).

\begin{figure}[hbt]
  \centering
  \includegraphics[width=\linewidth]{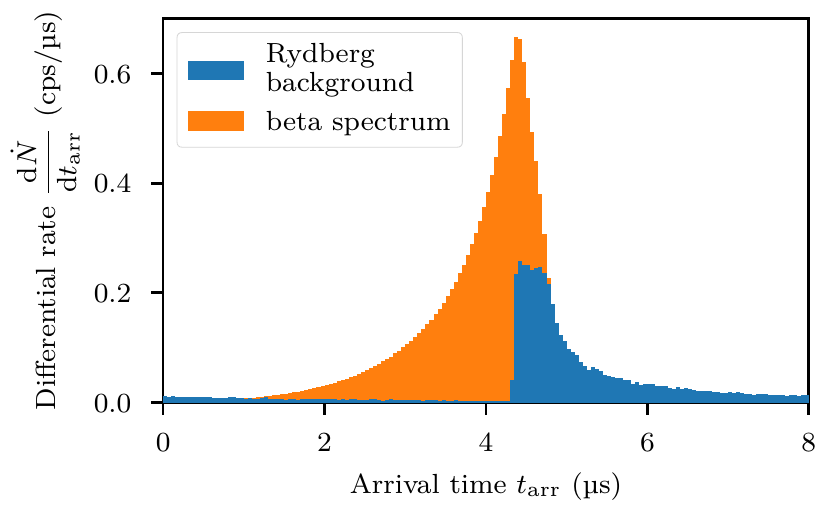} 
  \caption{TOA spectrum of ``Rydberg background'' electrons (blue) in comparison with signal electrons from the beta spectrum (orange) for a retarding potential $E_0-qU=11.3~\up{eV}$ below the endpoint. The ``Rydberg background'' has a total rate of 261~mcps, while the beta electron rate is 766~mcps.}
  \label{fig:toa_rydberg}
\end{figure}

Thus, the ``Rydberg background'' electrons can partly be separated from signal electrons because of their distinct contribution to the overall time-of-arrival spectrum.
For simplicity, we assume in our toy model that the total spectrometer background rate of \SI{293}{mcps} is described mostly by this Rydberg scenario, and a small contribution of about 32~mcps caused by radioactivity nearby the detector or other sub-dominant background sources.
Figure \ref{fig:toa_rydberg} demonstrates that the tfTOF mode separates the ``Rydberg background'' electrons from the signal electrons such that even with the much higher background rate from the spectrometer of \SI{261}{\mcps} -- the sensitivity deteriorates only slightly to
$\sigma^\up{tfTOF}_\up{stat}(m^2(\nu_e))=$ \SI{0.015}{\electronvolt\squared}, see tab.  \ref{tab:results}.

In this study we use the same waveform (given by eq.~(\ref{eq:superposition}) and eq.~(\ref{eq:modulation})) for the potential ramping in the tfTOF section for every retarding voltage $U_\up{MS}$ applied to the main spectrometer.
This leaves some room for further optimization because an individual waveform could be used for each retarding potential.

\section{Conclusion}
\label{sec:conclusion}

The advantages of a MAC-E-filter -- large solid angle acceptance and high energy resolution -- are connected to its integrating characteristics, which also constitutes some disadvantage because a spectrum needs to be measured consecutively at many different retarding energies. Adding a differential method to the energy high-pass MAC-E-filter can significantly improve statistics, or lower the measurement time, because a smaller number of retarding potentials is needed.
The potential benefits of time-of-flight methods turning a MAC-E-filter into a differential mode have already been demonstrated in another study \cite{Steinbrink2013}, but they require knowledge of the start time of the electron (ion) under investigation. Unfortunately, the start time of the electron (ion) can not be determined in all cases without significantly modifying its energy information. 

In this study we investigated an alternative approach, which we call ``time-focusing time-of-flight'' (tfTOF), where the start time does not have to be known.
Applying a time dependent electric potential to a MAC-E-filter-like tfTOF section in the beamline allows one to focus the arrival time of particles of a certain energy.
Thus, energy information can be inferred from the recorded time-of-arrival spectrum.
Combining the tfTOF spectrometer with a standard MAC-E-filter in front of it allows for a quasi-differential measurement of a small part of a continuous spectrum, keeping the excellent solid angle acceptance and energy resolution of the MAC-E-Filter. The spectral fraction where the quasi-differential measurement is enabled by the tfTOF method is selected by the static electric potential of the standard MAC-E filter. The parameters of the time-dependent electric potential in the tfTOF section then determine which part of the remaining spectrum is time-focused.

In order to evaluate the achievable improvement of the new quasi-differential method we studied its application to the neutrino mass search with the KATRIN experiment.
We found that we could improve the statistical sensitivity to $m^2(\nu_e)$ by a factor of two if we assume the same background as in the Design Report \cite{KATRIN2005}. This corresponds to an improvement of the sensitivity to the neutrino mass $m(\nu_e)$ of  a factor $\sqrt{2}$ which might be too small to justify the huge effort for setting up an additional MAC-E-filter as a tfTOF section.

Unfortunately, the background rate of KATRIN is still much enhanced w.r.t. the value anticipated in the KATRIN Design Report \cite{KATRIN2005}. 
Since this enhanced ``Rydberg background'' exhibits special spectral and angular characteristics, the full power of tfTOF can be exploited in this case. 
A dedicated tfTOF section behind the KATRIN main spectrometer would allow to separate the KATRIN spectrometer background electrons from the signal electrons in the time-of-arrival spectrum (for the ``Rydberg background'' electron model). The tfTOF approach yields a slightly higher sensitivity -- although the background rate for the tfTOF case was assumed like in KATRIN's first science run to be a factor 30 higher  -- w.r.t. the standard integral KATRIN mode at the design background rate of \SI{10}{\mcps}.

While the sensitivity to $m^2(\nu_e)$ of our method still lacks behind the potential of a real TOF measurement requiring an ``electron tagger'' by a factor of 2.8, one has to remember that such a tagger device without too much disturbance of the electron energy or momentum has not been realized as of yet.

In the study summarized above, optimization was performed to achieve the best neutrino mass sensitivity at low background rates.
This optimization already let the tfTOF method become much less sensitive to ``Rydberg background'' electrons.
Further optimization of the parameters of the electric potential of the tfTOF section could directly address the spectral and angular distribution of this specific background.
It is possible that focusing the optimization on the background will allow to achieve similar improvements with a tfTOF section shorter in length and smaller in diameter than the KATRIN main spectrometer. 
The study of a possible application with KATRIN demonstrates that the tfTOF method can be used in general, and in particular to modify the sensitivity to certain energy and angular characteristics of an electron or ion source.

Furthermore, the method still has some room for optimization regarding the choice of retarding potentials and the parameters of the focusing potential.
Utilizing additional Monte Carlo studies along with already planned experimental measurements with a pulsed electron source with known start time, an optimized waveform might be derived.

\begin{acknowledgements}
This project was funded by BMBF under contract number 05A17PM3.
\end{acknowledgements}

\bibliographystyle{spphys}
\bibliography{references}

\end{document}